\documentclass[aip, reprint, nofootinbib, longbibliography]{revtex4-1}
\usepackage{graphicx}
\usepackage{glossaries}
\usepackage{amsmath}
\usepackage{amssymb,amsmath}
\usepackage[usenames,dvipsnames]{color}
\usepackage[margin=.57in]{geometry}
\usepackage{braket}
\usepackage{enumitem}
\usepackage{mathtools}

\usepackage[english]{babel} 
 \usepackage{underoverlap}
\usepackage{ntheorem}
\theoremseparator{:}
\usepackage{ulem}

\usepackage{url}

\usepackage{filecontents}
\pdfoutput=1

\begin{document}
\normalem

\title{Crystallization in Three-Dimensions: Defect-Driven Topological Ordering and the Role of Geometrical Frustration}

\author{Caroline S. Gorham}
\email{caroling@cmu.edu}
\affiliation{Department of Materials Science and Engineering, Carnegie Mellon University, Pittsburgh, PA 15213, USA}

\author{David E. Laughlin}
\email{laughlin@cmu.edu}
\affiliation{Department of Materials Science and Engineering, Carnegie Mellon University, Pittsburgh, PA 15213, USA}

\begin{abstract}
Herein, fundamentals of topology and symmetry breaking are used to understand crystallization and geometrical frustration in topologically close-packed structures. This frames solidification from a new perspective that is unique from thermodynamic discussions. Crystallization is considered as developing from undercooled liquids, in which orientational order is characterized by a surface of a sphere in four-dimensions (quaternion) with the \emph{binary polyhedral} representation of the preferred orientational order of atomic clustering inscribed on its surface. As a consequence of the dimensionality of the quaternion orientational order parameter, crystallization is seen as occurring in ``restricted dimensions.'' Homotopy theory is used to classify all topologically stable defects, and third homotopy group defect elements are considered to be generalized vortices that are available in superfluid ordered systems. This topological perspective approaches the liquid-to-crystalline solid transition in three-dimensions from fundamental concepts of: Bose-Einstein condensation, the Mermin-Wagner theorem and Berezinskii-Kosterlitz-Thouless (BKT) topological-ordering transitions. In doing so, in this article, concepts that apply to superfluidity in ``restricted dimensions'' are generalized in order to consider the solidification of crystalline solid states.
\end{abstract}
\maketitle

\section{Introduction}

This article aims to elucidate the topological-ordering mechanisms that lead to solidification of undercooled atomic liquids into crystalline ground states. A topological viewpoint is pursued on the liquid-to-solid transition, which unifies several perspectives. This topological viewpoint on crystallization makes use of a quaternion orientational order parameter, and is thereby a generalization of Bose-Einstein condensation phenomena of superfluids (complex) in ``restricted dimensions'' (Mermin-Wagner theorem~\cite{mermin_absence_1966}). {This perspective builds on our recent computational work, that elucidates the topological origins of orientational ordering phase transitions in 4D quaternion ordered systems~\cite{gorham_su2_2018}, that is an extension of the classical 2D XY model~\cite{kosterlitz_critical_1974}. }{This generalization is related to the extension of quantum Hall effect phenomena (topological-order) from 2D to 4D, by considering their mathematical frameworks within the complex and quaternion compact (gauge) Lie algebra domains~\cite{zhang_four-dimensional_2001, bernevig_eight-dimensional_2003, lohse_exploring_2018}.} 

General features of Bose-Einstein condensation of superfluids (complex), for which the free energy function is a `Mexican hat' and vortices are available as topological defects, are discussed in Section~\ref{sec:ssb_complex}. In particular, spontaneous symmetry breaking (SSB) in ``bulk dimensions'' is distinguished from topological-ordering that occurs for complex ordered systems that exist in ``restricted dimensions.'' Furthermore, the manifestation of a frustration-induced quantum phase transition that occurs for complex ordered systems in ``restricted dimensions'' is introduced in Section~\ref{sec:ssb_complex}. In Section~\ref{sec:unwrapping_disclinations}, the concept of Bose-Einstein condensation is generalized beyond complex ordered systems by making use of a quaternion orientational order parameter to characterize the degree of order in undercooled atomic liquids. The notions of Berezinskii-Kosterlitz-Thouless topological ordering transitions~\cite{berezinskii_destruction_1971, kosterlitz_ordering_1973} in ``restricted dimensions,'' are therein generalized beyond superfluids (complex) to consider the formation of crystalline solid states (quaternion) including geometrically frustrated topologically close-packed (TCP) structures~\cite{frank_complex_1958, frank_complex_1959} (e.g., Frank-Kasper structures).

\section{Spontaneous Symmetry Breaking and ordering in ``bulk'' and ``restricted dimensions'' for superfluids (complex)}
\label{sec:ssb_complex}

\begin{figure*}
\centering
\includegraphics[scale=1]{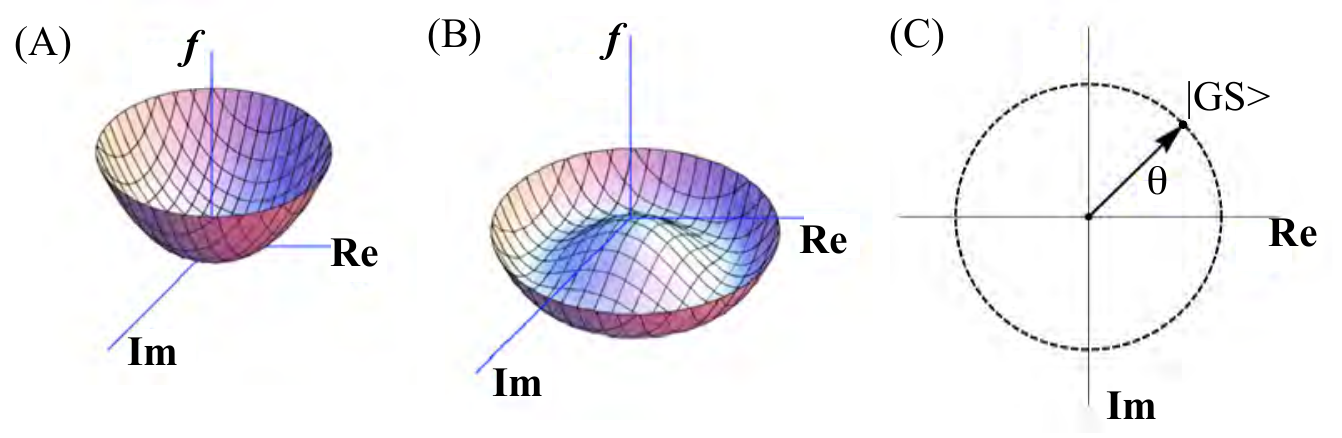}
\caption{Superfluid free energy. (A,B) The vertical axis represents the superfluid free energy, $f = \alpha |\psi|^2 + \beta|\psi|^4$, for a complex order parameter field $\psi=|\psi|e^{i\theta}$. The horizontal axes are the real and imaginary parts of $\psi$ in the complex plane. (A) For $T>T_\text{BEC}$, $\alpha>0$, $\beta>0$ and, the $U(1)$ symmetry group of a high-temperature fluid is unbroken and the free energy is at a minimum for $\psi=0$. The free energy is invariant to complex rotation $\psi\rightarrow \psi e^{i\theta}$, and may therefore be rotated about the vertical axis for $\theta \in [0,2\pi]$. (B) For $T<T_\text{BEC}$, $\alpha<0$, $\beta>0$ and, the superfluid spontaneously breaks the $U(1)$ symmetry at the origin ($\psi=0$) by adopting a particular ground state on $\mathcal{M}=S^1$. The complete manifold of degenerate ground states has $O(2)$ symmetry, and is known as the `Mexican hat.' (C) Schematic of the ground state manifold of superfluids, $\mathcal{M}=S^1$ of radius $|\psi|=\sqrt{\alpha/2\beta}$ and $\theta\in[0,2\pi]$. }
\label{fig:SSB}
\end{figure*}

A well-known example of SSB is the formation of superfluid Bose-Einstein condensates of electrons or helium particles, by broken $U(1)$ (complex) symmetry, in three-dimensions. At high-temperatures, above the Bose-Einstein condensation temperature ($T_\text{BEC}$), the system is disordered/normal and the free energy is minimized at the origin of the complex plane (Fig.~\ref{fig:SSB} A). This free energy is invariant under the symmetry of complex rotations $\psi\rightarrow e^{i\theta}\psi$, where $\theta\in[0,2\pi]$. SSB is only possible for complex ordered systems that exist in three-dimensions because this dimension plays the role of a  ``bulk dimension'' for the complex Lie algebra domain. Below the superfluid transition temperature $T_\text{BEC}$, the free energy that describes the superfluid state is minimized for a finite amplitude of a complex order parameter and for any particular ground state (Fig.~\ref{fig:SSB} B and C) on $\mathcal{M}=S^1$ that applies globally.

In contrast to SSB, in ``bulk dimensions,'' systems that exist in ``restricted dimensions'' are prevented from undergoing a conventional disorder-order phase transition (by SSB) at finite temperatures (Mermin-Wagner theorem~\cite{mermin_absence_1966}). In complex ordered systems (i.e., superfluids), 2D and 1D  are considered to be ``restricted dimensions'' as a consequence of the abundance of misorientational fluctuations (in $\theta$) that develops just below $T_\text{BEC}$ and prevents phase-coherency. In these scenarios, a finite amount of undercooling (below $T_\text{BEC}$) is required prior to the formation of a phase-coherent superfluid state at low-temperatures. 

Phase-coherent superfluid states, that develop in ``restricted dimensions,'' do so by a Berezinskii-Kosterlitz-Thouless (BKT) type topological ordering transition within the gas of misorientational fluctuations that take the form of vortex point defects and anti-defects. In such systems, since both the energy and entropy of isolated point defects depends logarithmically on the system size, energy dominates the thermodynamics at low-temperatures and these defects will bind into low-energy (sum-0) pairs at the BKT transition temperature. This topological ordering event enables the existence of a superfluid ground state (phase-coherent) in ``restricted dimensions.''

Superfluid ordered systems that exist in ``restricted dimensions'' are well-modeled mathematically using $O(2)$ quantum rotor models~\cite{sachdev_quantum_2011}, more commonly known as Bose-Hubbard models~\cite{herbut_dual_1998, freericks_phase_1994}, on $D-$dimensional lattices (for $D\leq 2$). These models allow for the manifestation of frustrated ground states, as a function of the ratio between the interaction strength ($J$) to the hopping amplitude ($t$) that describes the mobility of bosons in the system. At zero temperature, in the thermodynamic limit, the system can be tuned between phase-coherent superfluid and phase-incoherent insulator states that are connected by a quantum phase transition~\cite{sachdev_quantum_2011}. 

These low-temperature states are achieved by dual Berezinskii-Kosterlitz-Thouless transitions~\cite{fazio_charge-vortex_1992, fazio_charges_2013}, of condensed particle and topological defect degrees of freedom. For example, in the laboratory, the application of a perpendicular magnetic field acts to a charged superfluid (superconductivity) acts to explicitly drive an asymmetry in the concentrations of magnetic vortices. Those with a sign corresponding to the direction of the external field become dominant~\cite{teitel_josephson-junction_1983, gantmakher_superconductor-insulator_2010}; with critical applied magnetic field, a quantum phase transition can be achieved and the phase-coherent superfluid is destroyed.

In Section~\ref{sec:unwrapping_disclinations}, the formation of crystalline ground states from clustered undercooled atomic fluids is discussed. In three-dimensions, a \emph{quaternion orientational order parameter} characterizes the degree of atomic clustering in the undercooled fluid such that crystallization is considered to be a direct higher-dimensional analogy to the formation of superfluids in ``restricted dimensions.''

\section{Formation of crystalline solids from undercooled atomic fluids in ``restricted dimensions''}
\label{sec:unwrapping_disclinations}

As a higher-dimensional realization of orientational ordering in ``restricted dimensions'' (Mermin-Wagner), consider the process of crystallization in three-dimensions. Above the melting temperature $T_M$, unclustered particles rotate fully in three-dimensions and the full orientational symmetry group is $G=SO(3)$. In considering the structure of topological defects throughout the system (by homotopy theory~\cite{mermin_homotopy_1978, mermin_topological_1979}), it is important to make use of the \emph{universal covering space}\footnote{A universal covering space $C$ of a space $G$ is a covering space that is simply connected, i.e., $\pi_1(C)=0$ while  $\pi_1(G)\neq0$.} of $SO(3)$ which is $SU(2)$. The unitary group of degree two has the topology of a spherical surface in four-dimensions (i.e., $S^3\in\mathbb{R}^4$). Unit quaternions provide the group structure for $SU(2)\cong S^3$, just as unit complex numbers do for $U(1)\cong S^1$ (see Figure 2).

{On SSB of a compact Lie group~\cite{goldstone_broken_1962,nambu_quasi-particles_1960}, i.e., complex $(\mathbb{C})$, quaternion ($\mathbb{H}$) or octonion ($\mathbb{O}$), a set of Goldstone modes are anticipated for each broken (imaginary) generator that defines the compact (gauge) group~\cite{goldstone_broken_1962,nambu_quasi-particles_1960}.} Just as a single Goldstone mode (phonon) is anticipated on spontaneous symmetry breakdown of the $U(1)$ group (superfluids), a set of three Goldstone modes (phonons) are anticipated on the breakdown of $SU(2)$ symmetry. In superfluid states, of broken $U(1)$ symmetry, this single long-wavelength Goldstone mode is responsible for the ``second sound'' that allows for heat transfer~\cite{kapitza_heat_1941, landau_theory_1941, kapitza_heat_1941, schmitt_superfluidity_2015, hu_second_2010}. Likewise, it is anticipated that the three phonon modes responsible for heat transfer in three-dimensional crystalline solids (1L, 2T) are the anticipated Goldstone modes on the breakdown of $SU(2)$ symmetry.

     \begin{figure}[t!]
  \centering
\includegraphics[scale=.6]{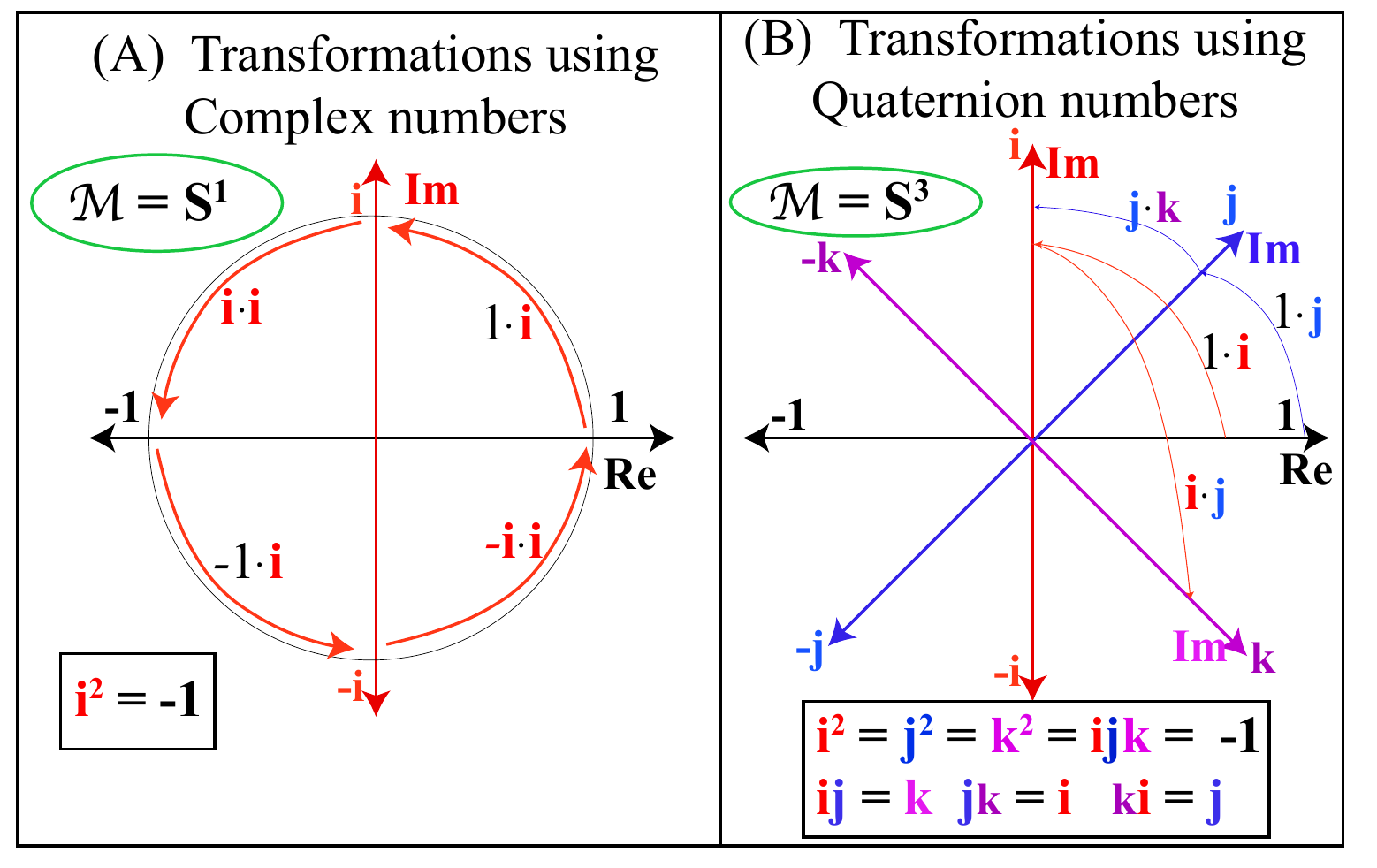}
\caption{(A) The product of complex numbers ($\mathbb{C}$) may be seen as $90^\circ$-rotations that span the \emph{complex plane} with the 2D basis of $\{1,\hat{i}\}$. Unlike real numbers ($\mathbb{R}$) that are not compact, complex numbers are compact under standard operations (by including $\sqrt{-1}$). Complex numbers then take the form: $z=x+y\sqrt{-1}$ where $x$ and $y$ are real numbers, and $\hat{i}^2=-1$ is an imaginary unit. When $x^2+y^2=1$, a complex number lies on the unit circle. (B) The next higher-dimensional algebra domain, forming a compact group, are the quaternion numbers ($\mathbb{H}$). The product of quaternion numbers may be seen as $90^\circ$-rotations in the \emph{quaternion plane} that spans the 4D basis $\{1,\hat{i}, \hat{j}, \hat{k}\}$. Just as complex numbers are formed as pairs of real numbers and an imaginary unit ($\hat{i}$), quaternion numbers may be formed in a similar manner as pairs of complex numbers: $q = (x_1 + y_1\hat{i}) + (x_2 + y_2\hat{i}) \hat{j}$ where $\hat{i}\hat{j}=\hat{k}$ and $\hat{i}^2=\hat{j}^2=\hat{k}^2=\hat{i}\hat{j}\hat{k}=-1$. When $x_1^2+y_1^2+x_2^2+y_2^2=1$, a quaternion number lies on the unit hypersphere ($S^3$).  While complex multiplication is Abelian, the order of operations of quaternion multiplication is important and hence the group is non-Abelian.}
\label{fig:transformations}
\end{figure}

{Furthermore, the number of imaginary generators that define the compact group (complex or quaternion) elucidates the available topological defects in complex or quaternion ordered systems. Just as superfluids of broken $U(1)$ symmetry permit $\pi_1(S^1)$ vortices  (that concentrate external complex rotational fields), quaternion ordered systems of broken $SU(2)$ symmetry permit $\pi_3(S^3)$ topological defects (that concentrate external quaternion rotational fields).} Just as vortices are spontaneously generated points in two-dimensions that act to prevent the development of conventional orientational order at finite temperatures (Mermin-Wagner theorem), $\pi_3(S^3)$ topological defects are spontaneously generated points in four-dimensions that act to prevent a conventional disorder-order transition~\cite{gorham_su2_2018, gorham_topological_2019} at finite temperatures. Undercooling below the melting temperature is therefore analogous to the prevention of SSB at finite temperatures for complex ordered systems that exist in ``restricted dimensions.''

Crystallization may occur after a finite amount of undercooling, as the clustered atomic fluid selects a particular orientational ground state from among the set of degenerate ground states on $\mathcal{M}$. That is, translational order obtained on the formation of a crystalline lattice (after a finite amount of undercooling) is coincident with the development of global orientational order {(quaternion)}. At the crystallization transition, by the adoption of an orientationally-ordered ground state, the topology of the relevant order parameter space changes from a sphere to a torus. Specifically, in spatial dimensions $D\geq 2$, the spherical orientational order parameter space ($\mathcal{M}$) of the undercooled liquid changes discretely at the crystallization transition by ``adding a handle'' to its surface~\cite{coleman_aspects_1988, zeeman_introduction_1966, tucker_branched_1936, mandelbaum_four-dimensional_1980, hoffman_adding_1993}. This forces the $D-$dimensional spherical order parameter space to decompose into the Cartesian product of $D$-spheres:
\begin{equation}
\mathcal{M}:S^D\rightarrow T^D   \cong \underbrace{S^1\times ... \times S^1}_{D},
\label{eqn:lattice_torus_M}
\end{equation} 
where $ T^D$ is the $D-$dimensional torus. 

\begin{figure}[t!]
  \centering
\includegraphics[scale=.4]{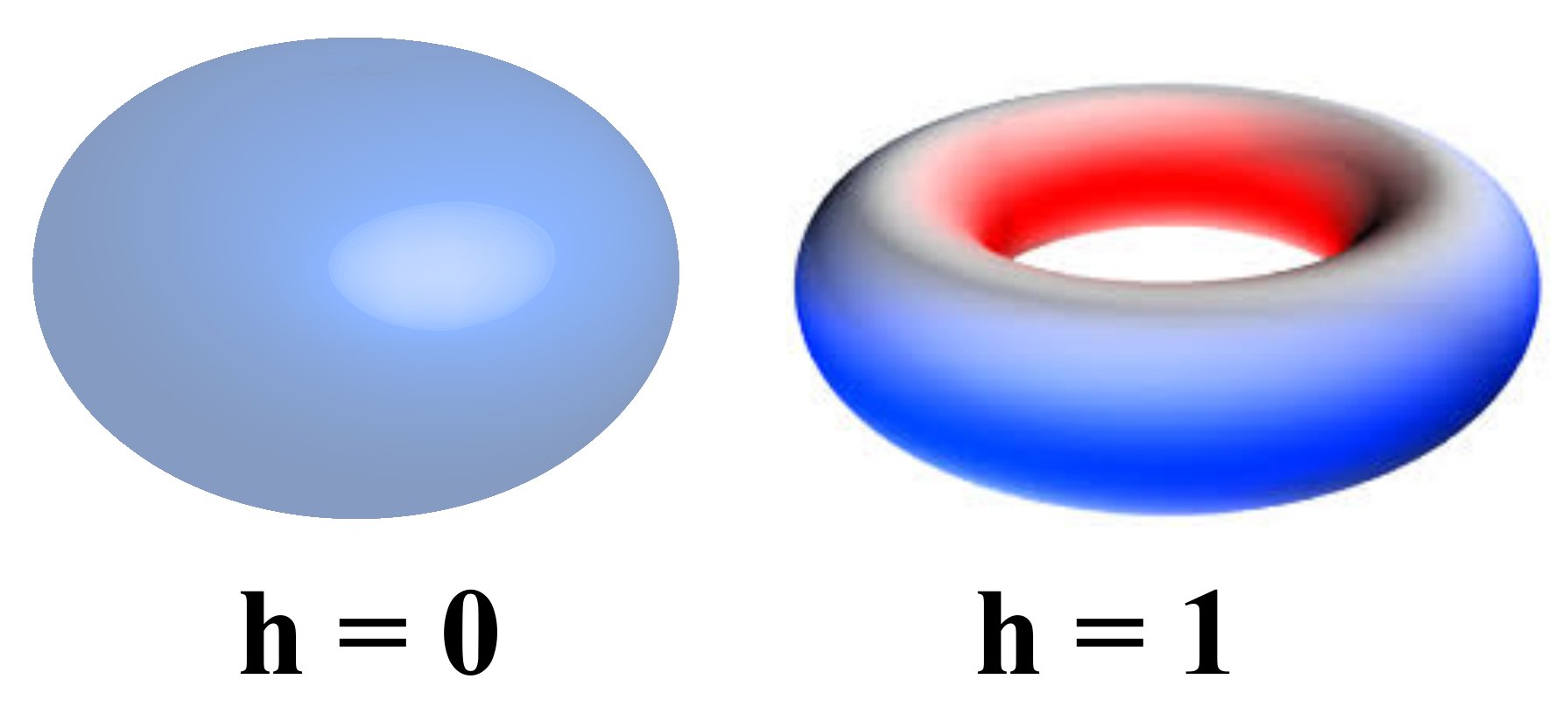}
\caption{In dimensions $D\geq 2$, sphere and tori are topologically distinct. The genus ($h$) of a surface $\mathcal{M}$ counts the number of ways that it can be cut into slices without it falling apart. A sphere has zero genus, and a torus has a singular genus. Hence, spheres are \emph{simply connected} and tori are not simply connected (i.e., $\pi_1(S^D)=0$ and $\pi_1(T^D)\neq 0$). While spherical surfaces have positive curvature everywhere, tori have regions of different curvature: sphere-like regions of positive curvature (blue), near the hole there is a saddle that has negative curvature (red), and on the top/bottom circles the local curvature is zero (grey).}
\label{fig:genus}
\end{figure}

Mechanisms of orientational ordering that drive the formation of crystalline solids may be elucidated by considering the discrete change in topology of $\mathcal{M}$ at the crystallization transition (Ref.~\onlinecite{gorham_topological_2019}). The change in topology of the order parameter manifold at the crystallization transition, as the genus topological invariant changes from zero to one, implies a change in the structure of the fundamental homotopy group~\cite{mermin_homotopy_1978, mermin_topological_1979} of topological defects (Figure~\ref{fig:genus}).

The genus topological invariant comes from the integration of \emph{curvature} over the surface, and is quantified directly via the Gauss-Bonnet theorem~\cite{allendoerfer_gauss-bonnet_1943, chern_simple_1944}. Spherical surfaces have positive integral Gaussian curvature, and toroidal surfaces have zero integral Gaussian curvature. The topology of the order parameter manifold that applies to the undercooled atomic fluid changes discretely to a torus (Eqn.~\ref{eqn:lattice_torus_M}) on the formation of a crystalline lattice. Thus, the development of translational order upon crystallization is (in essence) a flattening of atomic vertices. 

In light of this, it is not surprising that many authors have made use of a flattening method to model the crystallization process~\cite{sadoc_order_1982, nelson_symmetry_1984, sachdev_theory_1984, sethna_relieving_1983, mosseri_polytopes_1985, sadoc_geometrical_2006, mosseri_geometrical_2008}. The flattening method that is most commonly used is an extension of the standard Volterra process~\cite{hull_introduction_2001}, known as the ``iterative flattening method.'' By the ``iterative flattening method,'' the curved order parameter space that is tessellated by preferred short-range orientational order~\cite{nelson_symmetry_1984} is unwrapped into an infinite tiling of Euclidean space. {The discrete change in topology of the order parameter space upon crystallization ensures that the transition is discontinuous (first order).}

     \begin{figure}[b!]
  \centering
\includegraphics[scale=.41]{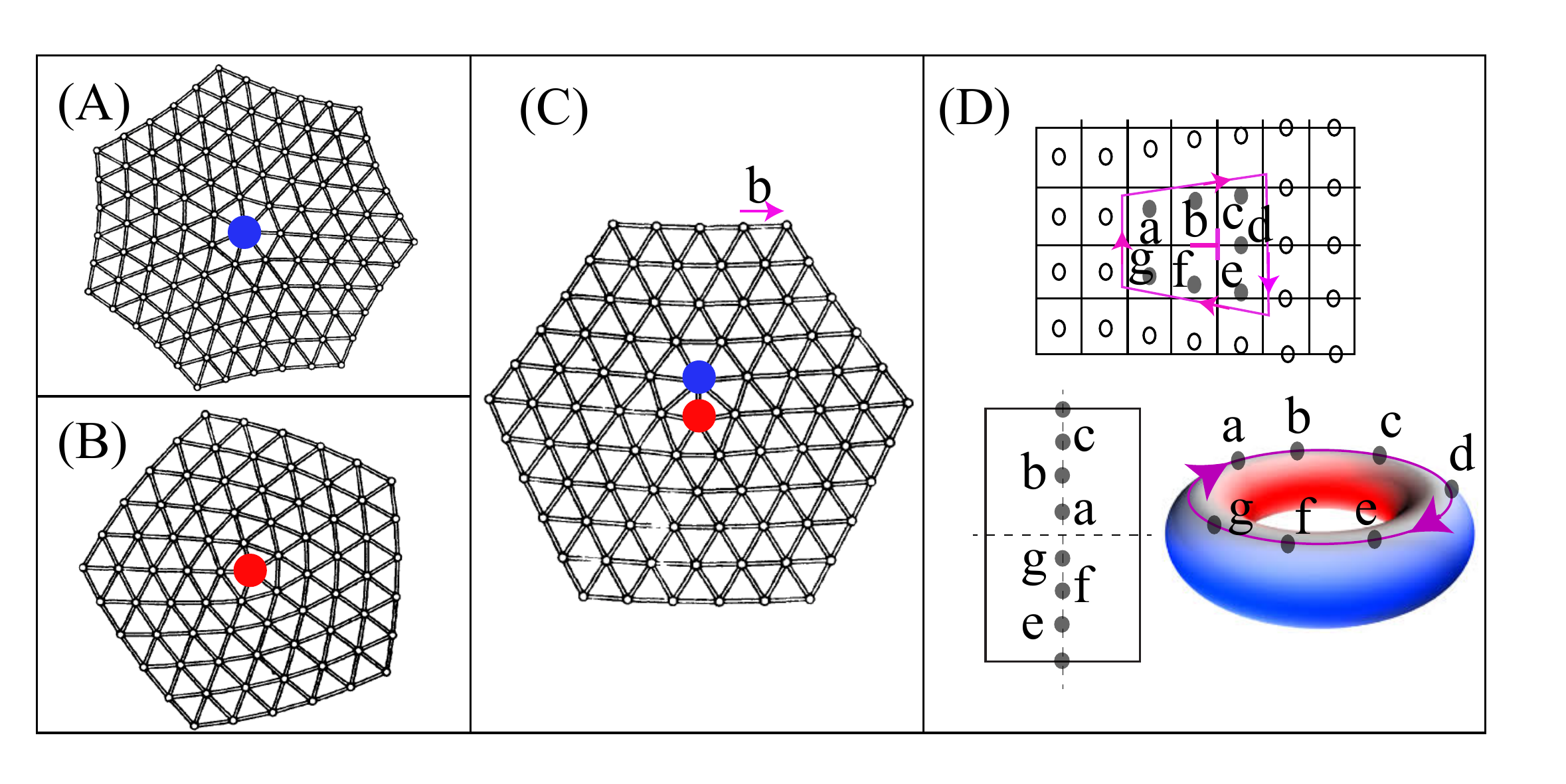}
\caption{(A) Negative and (B) positive wedge disclination point defects in two-dimensions.  (C) Pairs of complementary wedge disclinations are edge dislocations, which are translational defects in the field of atomic displacements of a crystalline lattice. [$A-C$ taken from Refs.~\onlinecite{pretko_fracton-elasticity_2018}] (D) Positions of atoms drift with respect to their ideal lattice positions as a closed loop is traversed around a dislocation. This corresponds to a loop around the central hole of the toroidal order parameter space, for atomic displacements in crystalline lattices. Rearranging the atoms slightly deforms the loop, but does not change the number of times it wraps around the torus (i.e., Burger's vector).}
\label{fig:disclinationpair}
\end{figure}

In addition to third homotopy group defects, due to atomic clustering in undercooled atomic fluids, wedge disclinations are also spontaneously generated at finite temperatures and act to prevent the formation of a crystalline solid state at the melting temperature. Ultimately, in order to achieve a crystalline solid at finite temperatures, a topological ordering event within this gas of wedge disclinations is required. This topological ordering event forces complementary wedge disclinations (Fig.~\ref{fig:disclinationpair} A and B) to bind into pairs that are considered to be dislocations~\cite{halperin_theory_1978, chaikin_principles_2000, pretko_fracton-elasticity_2018, yazyev_polycrystalline_2014} (Fig.~\ref{fig:disclinationpair} C and D) that are excitations from the crystalline ground state. This concept was originally discussed for 2D crystallization by Halperin and Nelson (Ref.~\onlinecite{halperin_theory_1978}), and is reminiscent of orientational ordering in 2D XY models~\cite{berezinskii_destruction_1971, kosterlitz_ordering_1973, kosterlitz_critical_1974}. In the absence of geometrical frustration, every lattice site in the crystalline ground state is equivalent and it is free of all topological defects. This is possible because, the plasma of topological defects that develops just below the melting temperature is \emph{balanced} between the concentrations of topological defects with equal and opposite signs (Fig.~\ref{fig:disclinationpair}A and B). In Sections~\ref{sec:crystallization_2d} and~\ref{sec:crystallization_3d}, crystallization phase transitions in 2D and 3D are considered from a topological viewpoint in the absence and presence of geometrical frustration.

\subsection{Crystallization in two-dimensions}
\label{sec:crystallization_2d}

For purposes of illustration, consider the crystallization process in 2D using the ``iterative flattening method.'' For two-dimensional solidifying fluids, below the melting temperature, the degree of orientational order is related to the regular tessellation on the surface of a sphere in three-dimensions ($S^2\in\mathbb{R}^3$) by the preferred orientational order of atomic clustering. Hence, regular polyhedra that tessellate $S^2\in\mathbb{R}^3$ act as ideal curved space models of orientational order in 2D undercooled atomic fluids. Only two of the five regular polyhedra, cubes ($\{4,3\}\in\mathbb{S}^2$) and tetrahedra ($\{3,3\}\in\mathbb{S}^2$), are developable\footnote{Developable curved surfaces may be flattened into infinite tilings without distortion or the introduction of permanent defects.} in flat (Euclidean) plane, these periodic nets are discussed in Section~\ref{sec:no_geometrical2D}. Crystallization of short-range orientational order that is not developable in flat space, i.e., that is considered to be geometrically frustrated, will be discussed in Section~\ref{sec:yes_geometrical2D}.

\subsubsection{Absence of geometrical frustration}
\label{sec:no_geometrical2D}

\begin{figure}[t!]
  \centering
\includegraphics[scale=.4]{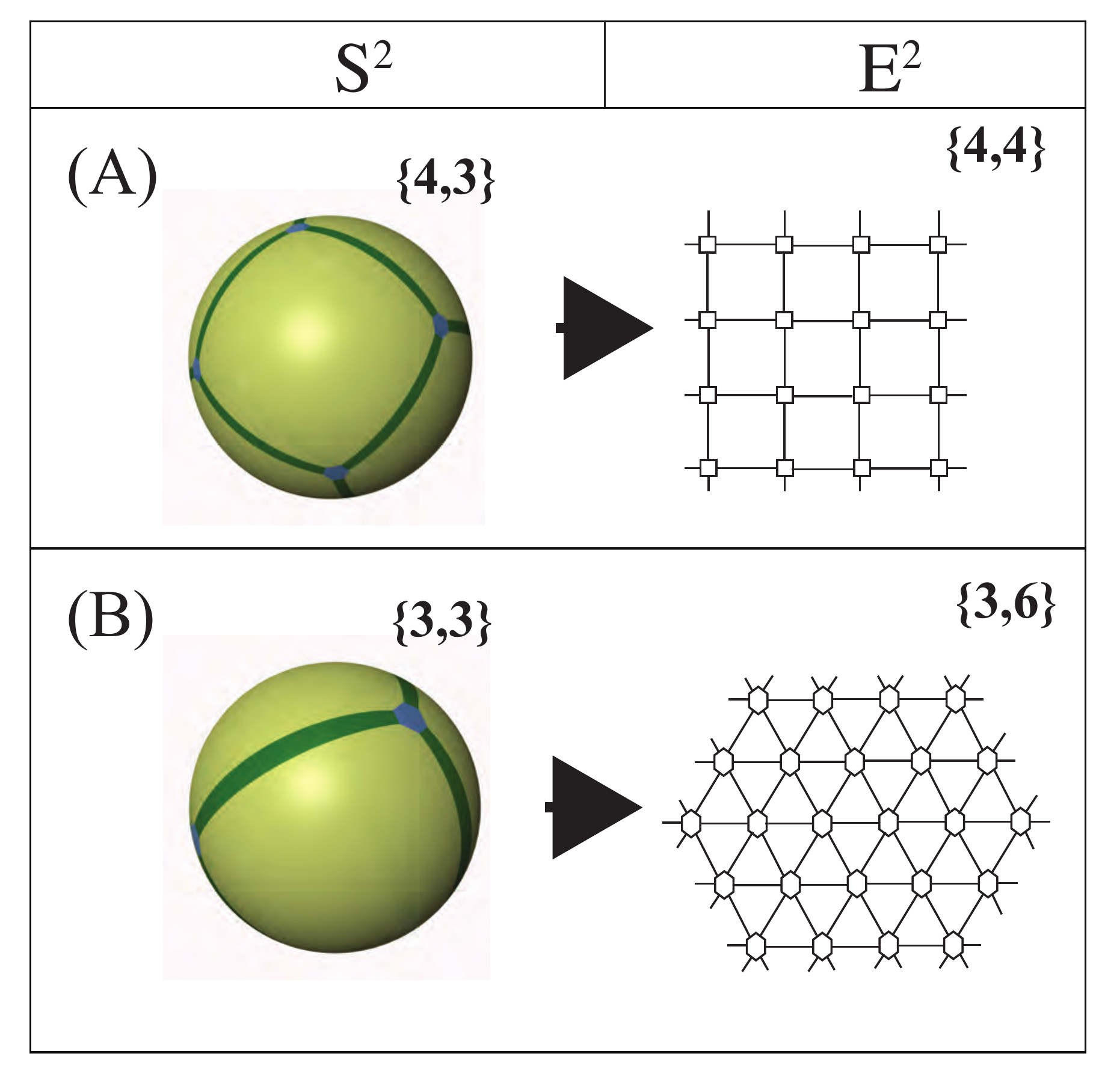}
\caption{Regular tilings of the plane can be made by: (A) unwrapping the faces of the cubic tessellation of $\{4,3\}\in S^2$ to form $\{4,4\}\in E^2$ and,  (B) developing $\{3,3\}\in S^2$ to generate $\{3,6\}$ nets. [Built using open-source KaleidoTile software (Ref.~\onlinecite{weeks_kaleidotile_nodate})]}
\label{fig:frustrated_plaquette}
\end{figure}

The two regular periodic tilings of the Euclidean plane are by squares and by equilateral triangles, which have the Schl\"{a}fli notation\footnote{Sch\"{a}fli notation defines~\cite{coxeter_regular_1973, de_graef_structure_2012} regular tessellations in arbitrary dimensions in the form $\{p,q,r,...\}$. In two-dimensions, $\{p,q\}$ gives a description of the system by accounting for the number of edges of each face ($p$) and the number of faces that meet at each vertex ($q$); in three-dimensions, the Scl\"{a}fli notation is $\{p,q,r \}$ and $r$ is the number of polyhedra cells per near-neighbor bond.} of $\{4,4\}$ and $\{3,6\}$ respectively. These 2D periodic nets can be generated by unwrapping the relevant developable curved space tessellations of vertices, that characterize orientational order in the undercooled fluid, onto the plane. In particular, square nets ($\{4,4\}$) can be generated by flattening cubic tessellations $\{4,3\}\in S^2$ onto the plane. In doing so, the coordination at each vertex increases from three to four so that square faces may generate infinite tilings (Fig.~\ref{fig:frustrated_plaquette} A). Similarly, on unwrapping tetrahedra $\{3,3\}\in\mathbb{S}^2$ each vertex becomes six-fold~\cite{sadoc_order_1982} in order to generate the $\{3,6\}$ infinite tiling (Fig.~\ref{fig:frustrated_plaquette} B). These periodic tilings are free of permanent disclination topological defects in the ground state, and the topological defects that act as excitations from the ground state are dislocations.

\subsubsection{Presence of geometrical frustration}
\label{sec:yes_geometrical2D}

\begin{figure}[t!]
  \centering
\includegraphics[scale=.15]{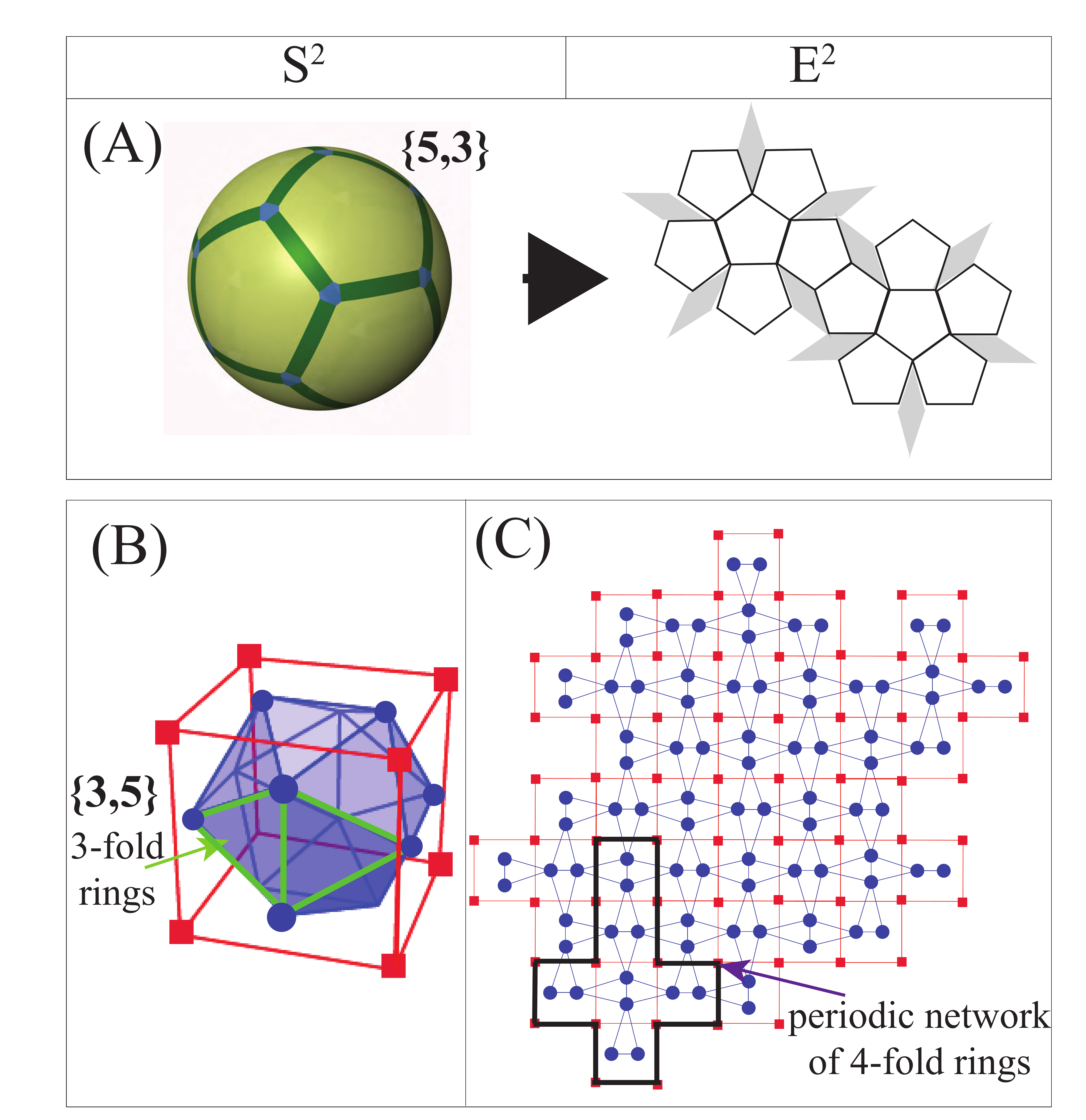}
\caption{(A) There is no regular pentagonal tiling of the plane because one cannot unwrap the dodecahedron $\{5,3\}\in S^2$ onto the plane without the introduction of permanent defects (grey). (B) Vertices of an icosahedron inscribed onto the faces of a cube. (C)  Tiling of the plane by the unwrapped cube with vertices of icosahedron inscribed is a 2D analogue of 3D Frank-Kasper structures. Permanent negative wedge disclinations are introduced at the vertices of the unwrapped cube (red squares), that correspond to the opening of some triangular rings of the icosahedron into four-membered rings~\cite{sadoc_order_1982}. [$B-C$ adapted from Ref.~\onlinecite{sadoc_order_1982}, ``Iterative flattening method.''] }
\label{fig:icosahedron_in_cube}
\end{figure}

Solidifying systems with five-fold symmetries are unable to generate infinite tilings of the Euclidean plane~\cite{penrose_pentaplexity_1979}, in contrast to systems with short-range orientational order that is developable in flat space. This is a consequence of geometrical frustration~\cite{sadoc_geometrical_2006}. For instance, the dodecahedron $\{5,3\}\in S^2$ can only be unwrapped onto the plane by the introduction of permanent defects (Fig.~\ref{fig:icosahedron_in_cube} A).

Ordered arrangements of topological defects in geometrically frustrated solids have been evidenced in physical (three-dimensional) crystalline systems, in cases of icosahedral short-range orientational order~\cite{frank_complex_1958, frank_complex_1959}. These crystalline solids, known as Frank-Kasper phases~\cite{frank_complex_1958, frank_complex_1959, doye_effect_1996}, are topologically close-packed (TCP) and the ordered network of permanent disclination lines that allows for their structural stability is known as the ``major skeleton'' network~\cite{de_graef_structure_2012}.  In order to generate a geometrically frustrated crystalline structure, using the ``iterative flattening method,'' the vertices of a particular orientational order parameter space are first inscribed onto the faces of a developable surface~\cite{sadoc_geometrical_2006} and then mapped into flat space. 

As a two-dimensional example using the ``iterative flattening method'' to generate a geometrically frustrated structure, consider the vertices of an icosahedron $\{3,5\}\in S^2$ inscribed onto the faces of a cube (Figure~\ref{fig:icosahedron_in_cube} B). Figure~\ref{fig:icosahedron_in_cube} C shows the infinite tiling of the plane made by the unwrapped the cell shown in Figure~\ref{fig:icosahedron_in_cube} B. This is a 2D analogue of 3D Frank-Kasper crystals.  In these cases of geometrical frustration, in order to ensure that the system is space-filling, permanent negative wedge disclinations are introduced at the vertices of the unwrapped cube (red squares), as an ordered ``major skeleton network.''

In the event of geometrical frustration, there is intrinsic positive curvature that is associated with the preferred short-range orientational order of atomic clustering. Physically, this intrinsic positive curvature (geometrical frustration) explicitly breaks the symmetry between the concentrations of disclinations in the undercooled liquid towards those that concentrate negative curvature at their core. This ensures that the overall Euclidean space remains flat. In scenarios of geometrical frustration~\cite{sadoc_order_1982, sadoc_geometrical_2006}, these excess negative wedge disclinations are unable to form bound states upon crystallization and will persist to the ground state as a periodic arrangement. These ordered networks of disclination topological defects in TCP structures are similar to Abrikosov vortex flux lattices that develop in frustrated thin-film superfluids~\cite{teitel_josephson-junction_1983, nelson_defects_2002}.

\subsection{Crystallization in three-dimensions}
\label{sec:crystallization_3d}

In three-dimensions, just below the melting temperature ($T_\text{M}$), the full orientational symmetry group of the high-temperature liquid ($G=SO(3)$) is broken locally by atomic clustering~\cite{frank_supercooling_1952, steinhardt_icosahedral_1981, nelson_liquids_1983}. The orientational order parameter space takes the form~\cite{mermin_homotopy_1978, mermin_topological_1979, nelson_liquids_1983}:
\begin{equation}
\mathcal{M}=\frac{SO(3)}{H} \equiv \frac{SU(2)}{H'},
\label{eqn:orderparam}
\end{equation} 
where $H \subset SO(3)$ and $H' \subset SU(2)$ is the \emph{binary polyhedral group}. {Because the group $SU(2)$ is isomorphic to the unit quaternions, local orientational ordering due to atomic clustering below the melting temperature may be characterized by the application of a \emph{quaternion orientational order parameter}~\cite{gorham_topological_2019}. The underlying space of the unit quaternions is the surface of a sphere in four-dimensions ($S^3\in \mathbb{R}^4$), which connects our proposed ordering field theory for crystallization with earlier models that consider solidification of geometrically frustrated orientational order using a regular ``crystal'' in curved space~\cite{nelson_liquids_1983, nelson_symmetry_1984, tarjus_frustration-based_2005} as a reference state.  

Topological stable defects, that can be drawn on the order parameter space (Eqn.~\ref{eqn:orderparam}), belong to the first and third homotopy groups~\cite{rivier_gauge_1990, mermin_homotopy_1978}:
\begin{equation}
\pi_1(\mathcal{M})={H'} , \,\,\,\, \,\,\,\,   \pi_3(\mathcal{M})=\mathbb{Z} ,
\label{eqn:topodefects}
\end{equation}
where $\mathbb{Z}=0,\pm1,\pm2...$ is a lattice of integers. In three-dimensional atomic systems~\cite{rivier_gauge_1990}, topological defects that belong to the first and third homotopy groups are: disclination lines and instantons; similarly, according to the topological charge equation~\cite{teo_topological_2010, teo_topological_2017}, these defects behave as planes and points in 4D/(3D+1t) quaternion ordered systems.} Owing to the existence of $\pi_3(\mathcal{M})$ defects as points in 4D/(3D+1t), solidifying atomic systems must be considered~\cite{gorham_su2_2018, gorham_topological_2019} to order in ``restricted dimensions.''  This is a direct extension of the Mermin-Wagner theorem~\cite{mermin_absence_1966}, that states that continuous symmetries cannot be spontaneously broken at finite temperatures in 2D/(1D+1t) systems. 

That is, just like in the $XY$ model and low-dimensional superfluids, there is no possibility for conventional long-range orientational order to develop in three-dimensional solidifying atomic systems at the melting temperature. Just below $T_M$, it is an abundance of misorientational fluctuations taking the form of a plasma of spontaneously generated topological defects (Eqn.~\ref{eqn:topodefects}) that prevents global orientational-order. Thus, undercooling is necessary as a consequence of the phase-destabilizing nature of spontaneously generated topological defects.

In order to allow for the development of crystalline solids at finite temperatures, a defect-driven Berezinskii-Kosterlitz-Thouless type transition~\cite{berezinskii_destruction_1971, kosterlitz_ordering_1973} must occur within this gas of topological defects that otherwise prevent the development of orientational order just below the melting temperature. Ultimately, it is the formation of bound pairs of third homotopy group defects/anti-defects and bound states of complimentary disclinations that drives crystallization~\cite{gorham_topological_2019} at a finite temperature below the melting temperature ($T_\text{defect-BKT}<T_M$). Three-dimensional crystallization, in the absence and presence of geometrical frustration, are discussed separately in Sections~\ref{sec:absence_3D} and Sections~\ref{sec:presence_3D}. Finally, an anticipated phase diagram for solidification in the vicinity of critical geometrical frustration (a ``self-dual critical point'') is presented in Section~\ref{sec:phase_diagram}.

\subsubsection{Absence of geometrical frustration}
\label{sec:absence_3D}

In the event that the relevant orientational order parameter space that characterizes atomic clustering (Eqn.~\ref{eqn:orderparam}) is developable in flat space, the crystalline solid state that forms below a critical temperature is free of permanent topological defects at zero Kelvin~\cite{nelson_symmetry_1984}. This is possible because, in the absence of geometrical frustration, the plasmas of misorientational topological defects that coexist with atomic clusters in the undercooled fluid are balanced such that there exists equal concentrations of topological defects with opposite signs. These balanced plasmas, of third and first homotopy group defects, can become entirely topologically ordered at low-enough temperatures~\cite{kosterlitz_ordering_1973, gorham_su2_2018, gorham_topological_2019}, by the formation of bound states via a defect-driven BKT-like mechanism. 

The formation of a perfect crystal is therefore akin to the formation of perfectly phase-coherent superfluid ground states in ``restricted dimensions,'' that are able to exist due to a prototypical BKT topological ordering transition within a balanced (classical) gas of vortices. Limiting ourselves to the discussion atomic clustering that has symmetries characterized by regular polytope, cubic crystal are the only space-filling regular tilings of 3D Euclidean space which derive from the hypercube $\{4,3,3\}$. These systems lack geometrical frustration and are hence flat everywhere. This is possible because no centers of concentrated curvature exist in the form of permanent topological defects.

\subsubsection{Presence of geometrical frustration}
\label{sec:presence_3D}

\begin{figure}[t!]
  \centering
\includegraphics[scale=.9]{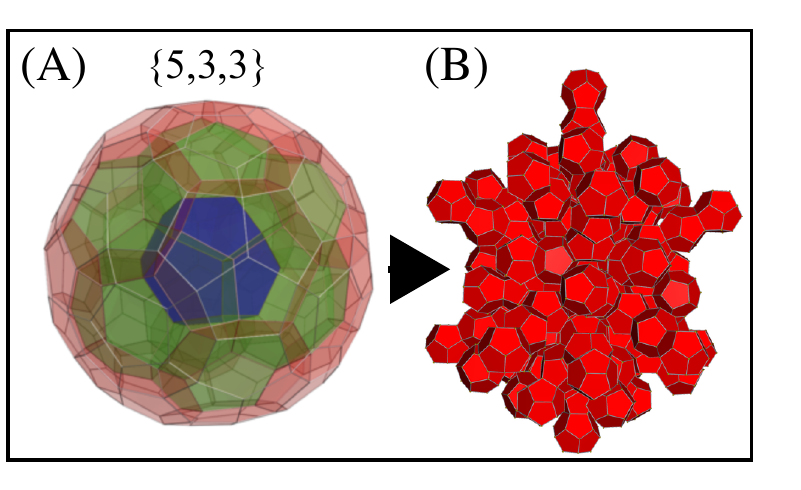}
\caption{(A) Perspective projection of the 120-cell polytope ($\{5,3,3\}\in S^3$). [Taken from Ref.~\onlinecite{teoh_4d_nodate}] (B) Net of $\{5,3,3\}\in S^3$ in Euclidean three-dimensional space. [Created with Stella4D software (Ref.~\onlinecite{webb_stella4D:_2003})].}
\label{fig:120-cell}
\end{figure}

In the presence of geometrical frustration, consider the flattening of the $\{5,3,3\}$ polytope (120-cell) into 3D Euclidean space (Fig.~\ref{fig:120-cell}) as a three-dimensional analogue to the flattening of the $\{5,3\}$ polyhedron into the plane (Fig.~\ref{fig:frustrated_plaquette}B). As a consequence of geometrical frustration, permanent topological defects are required in the flat space. Just as $\{5,3\}$ and $\{3,5\}$ are dual to one another, the dual of the  $\{5,3,3\}$ polytope is the $\{3,3,5\}$ polytope whose vertices are the 120-elements of the binary icosahedral group (i.e., the lift of $Y\in SO(3)$ into $SU(2)$). Thus, the $\{3,3,5\}$ polytope represents the orientational order parameter space of undercooled solidifying systems that exhibit an energetic preference for short-range icosahedral coordination of atomic clustering~\cite{frank_supercooling_1952, nelson_liquids_1983, nelson_symmetry_1984, widom_short_1988, venkataraman_curved_1985}. Such systems are geometrically frustrated, because, despite the realization of a regular ``crystal'' on the surface of the three-sphere in four-dimensions~\cite{coxeter_regular_1973, nelson_liquids_1983, nelson_symmetry_1984}. the preferred short-range orientational order (e.g., icosahedrally-coordinated systems) is unable to fill all of space (Fig.~\ref{fig:120-cell}). 

Geometrically frustrated crystalline solid states are known as topologically close-packed (TCP); Frank-Kasper structures~\cite{frank_complex_1959}, that express short-range icosahedral order, are a particular instance of TCP crystalline solids. Geometrical frustration is quantified as the curvature mismatch between the orientational order parameter space of the undercooled liquid $\mathcal{M}$ (Eqn.~\ref{eqn:orderparam}), and flat Euclidean space~\cite{nelson_symmetry_1984}. This curvature mismatch explicitly drives an asymmetry in the concentrations of positive and negative disclinations and third homotopy group defects (Eqn.~\ref{eqn:topodefects}), in order to ensure that the overall space remains flat on average~\cite{gorham_topological_2019}. In particular, excess negative wedge disclination lines neutralize remnant positive curvature that is attributed to atomic vertices that retain geometrically-frustrated coordination~\cite{sadoc_geometrical_2006}. This is in analogue to magnetic frustration of $O(2)$ Josephson junction arrays in the presence of an applied magnetic field~\cite{teitel_josephson-junction_1983, fazio_charge-vortex_1992}, which necessitates that the concentrations of magnetic vortices are shifted towards those that carry a sign corresponding to the direction of the external field~\cite{gantmakher_superconductor-insulator_2010}. 

In the presence of frustration, excess signed topological defects are unable to form bound pairs and persist to the ground state in a periodic manner~\cite{teitel_josephson-junction_1983}. This ensures an absence of configurational entropy at 0 K, in order to satisfy the third law of thermodynamics~\cite{wilks_third_1961, kittel_thermal_1970}. In analogue to frustrated ground states of $O(2)$ quantum rotor models~\cite{teitel_josephson-junction_1983}, geometrically-frustrated crystalline ground states no longer display perfect long-range orientational-order. Instead, the set of scalar phase angle parameters that characterize the orientational order parameter will vary from site to site to incorporate the \emph{major skeleton network} of topological defects~\cite{nelson_liquids_1983, nelson_symmetry_1984}. A periodic array of signed third homotopy group defects, that are points in (3D+1t) spacetime, also persist to the ground state of TCP structures; however, unlike disclination lines, these topological defects are not observable as structural constituents in three-dimensions unlike the ordered ``major skeleton network.''

\subsubsection{Anticipated phase diagram}
\label{sec:phase_diagram}

\begin{figure}[t!]
\centering
\includegraphics[scale=1]{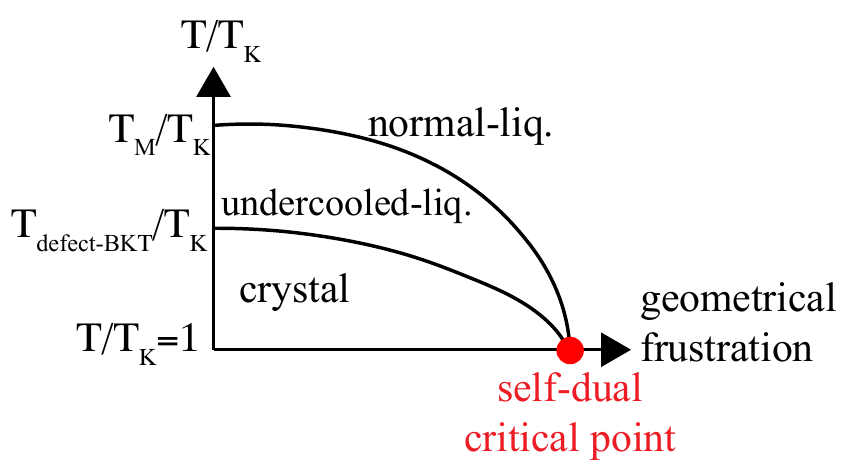}
\caption{Anticipated phase diagram for crystallization, in coordinates $T/T_\text{K}$ versus geometrical frustration where $T_\text{K}$ is the finite Kauzmann temperature~\cite{kauzmann_nature_1948, stillinger_kauzmann_2001}. A defect-driven BKT-like topological transition occurs at $T_\text{defect-BKT}$, that separates crystalline solid states from undercooled fluids that develop below the melting temperature ($T_M$). With increasing geometrical frustration, $T_\text{defect-BKT}$ is suppressed towards a minimum value (the Kauzmann temperature, $T_\text{K}$) that marks a first order transition between crystalline and non-crystalline solid states.}
\label{fig:curvature_sliding}
\end{figure}

Figure~\ref{fig:curvature_sliding} depicts a schematic of an anticipated phase diagram for three-dimensional crystallization~\cite{gorham_topological_2019}, in coordinates of reduced temperature versus geometrical frustration. A finite amount of undercooling below the melting temperature ($T_M$) is required, prior to crystallization. A defect-driven BKT transition marks crystallization  ($T_\text{defect-BKT}<T_M$), separating the solid state from the undercooled fluid. With increasing geometrical frustration, $T_\text{defect-BKT}$ is suppressed in the same way that the transition towards a phase-coherent superconducting state of $O(2)$ Josephson junction arrays is suppressed in the presence of a transverse magnetic field~\cite{fazio_charge-vortex_1992, teitel_josephson-junction_1983, gantmakher_superconductor-insulator_2010}. With increasing geometrical frustration, the distance between permanent topological defects in the ground state becomes minimized. 

Above a critical value of geometrical frustration, the concentrations of topological defects in the undercooled fluid become entirely biased such that bound states (e.g., dislocations) are unable to form and the ground state is no longer crystalline. In three-dimensions, the hypothetical solid state that forms at a critical value of geometrical frustration lacks translational order but retains short-range orientational order. It has been suggested previously that~\cite{gorham_topological_2019} the formation of a solid state at a critical geometrical frustration marks the so-called ``ideal'' glass transition that is referred to in the literature as the Kauzmann point~\cite{kauzmann_nature_1948}.

The transition between geometrically-frustrated solid states and non-crystalline solid states, that occurs at the Kauzmann point (``ideal'' glass transition), takes place at a finite temperature which corresponds to a first-order transition~\cite{baturina_superinsulatorsuperconductor_2013}. {Non-crystalline (orientationally-disordered) solid states are considered to be `dual' to crystalline solids, in which topological defects that persist to 0 K form a tangled arrangement. Unlike the zero configurational entropy at 0 K in TCP crystalline structures, this tangled arrangement of defects leads to finite residual configurational entropy.}

\section{Summary and conclusions}

Two fundamental concepts that drive the manifestation of a particular ordered state of matter are: the topology of the order parameter space and, the spatial dimensionality of the ordered system. In this article, we have discussed a particular example of the role of topology in condensed matter physics by approaching the crystallization process in three-dimensions using a \emph{quaternion orientational order parameter}. Owing to the fact that the solidifying system exists in ``restricted dimensions,'' conventional orientational order is prevented at the melting temperature and undercooling is viewed as. In light of this, we have suggested that the crystallization process in three-dimensions is a higher-dimensional realization of the formation of phase-coherent complex ordered systems in ``restricted dimensions,'' two- and one- (Mermin-Wagner theorem).

We have emphasized that, it is the discrete change in the topology of the order parameter space upon the formation of a crystalline lattice (by the addition of a handle to the otherwise spherical surface) that forces the emergence of bound states of topological defects via a defect-driven Berezinskii-Kosterlitz-Thouless like transition in three-dimensions. As a consequence of ordering in ``restricted dimensions,'' geometrical frustration is possible in three-dimensional crystalline solid states. Geometrical frustration forces a periodic arrangement of signed topological defects into crystalline ground states which are known as \emph{topologically close-packed} (TCP). Critical geometrical frustration entirely suppresses the crystalline ground state, at the Kauzmann point.

\section{Acknowledgements}
C.S.G. acknowledges support from NASA's Office of Graduate Research through the Space Technology and Research Fellowship (NSTRF). We also acknowledge support from the ALCOA Chair in Physical Metallurgy.

\bibliography{\jobname}

\end{document}